\documentclass[prl,twocolumn,showpacs,amsmath,amssymb,superscriptaddress,floatfix,nofootinbib,10pt]{revtex4}

\usepackage[utf8]{inputenc}
\usepackage{mathrsfs,bm}
\usepackage{longtable,lscape}
\usepackage{txfonts}
\usepackage{amssymb}
\usepackage{indentfirst}
\usepackage{graphicx,booktabs}
\usepackage{multirow}
\usepackage{overpic}
\usepackage{color}
\usepackage{amssymb}
\usepackage{hyperref}
\usepackage{ulem}
\usepackage{booktabs}

\begin{document}
\title{Prediction of an $\Omega_{bbb}\Omega_{bbb}$ dibaryon in  the extended one-boson exchange model}

\author{Ming-Zhu Liu}
\affiliation{School of Space and Environment, Beihang University, Beijing 102206, China}
\affiliation{School of Physics, Beihang University, Beijing 102206, China}

\author{Li-Sheng Geng}\email[Corresponding author:]{ lisheng.geng@buaa.edu.cn}
\affiliation{School of Physics, Beihang University, Beijing 102206, China}
\affiliation{
Beijing Key Laboratory of Advanced Nuclear Materials and Physics,
Beihang University, Beijing 102206, China}
\affiliation{School of Physics and Microelectronics, Zhengzhou University, Zhengzhou, Henan 450001, China}
\affiliation{Beijing Advanced Innovation Center for Big Data-Based Precision Medicine, School of Medicine and Engineering, Beihang University, Beijing, 100191}

\begin{abstract}
 Ever since Yukawa proposed that the pion is  responsible for mediating the nucleon-nucleon interaction, meson exchanges have been widely used in understanding hadron-hadron interactions. The most studied mesons are the $\sigma$, $\pi$, $\rho$, and $\omega$, while other heavier mesons  are often argued to be less relevant because they lead to short range interactions. However, the ranges of interactions should be compared with the size of the system under study  but not in absolute terms. In this work, we propose that one charmoninium exchange is responsible for the formation of the $\Omega_{ccc}\Omega_{ccc}$ dibaryon, recently predicted by lattice QCD simulations. The same approach can be extended to the strangeness and bottom sectors, leading to the prediction on the existence of $\Omega\Omega$ and $\Omega_{bbb}\Omega_{bbb}$ dibaryons, while the former is consistent with existing lattice QCD results, the latter remains to checked. In addition, we  show that the Coulomb interaction may break up the $\Omega_{ccc}\Omega_{ccc}$ pair but not  the  $\Omega_{bbb}\Omega_{bbb}$ and $\Omega\Omega$ dibaryons, particularly, the latter.

\end{abstract}


\maketitle

   \textit{Introduction:} In 1935, Yukawa proposed the one-pion exchange theory to explain the strong nuclear force~\cite{Yukawa:1935xg}. In the following years,  M. Taketani et al. employed one-pion  and two-pion exchange potentials to explain the then available nucleon-nucleon ($NN$) scattering data~\cite{Taketani:1935xg,Taketani:1952xg}.
    Later, it was realized that the exchange of heavier mesons, such as $\sigma$, $\rho$, and $\omega$, was needed to explain the rich experimental data, leading to the so-called  one-boson exchange  (OBE) model~\cite{Sawada:1962xg}, based on which  the Bonn group~\cite{Machleidt:1987hj} and Nijmegen group~\cite{Stoks:1994wp} constructed high-precision nuclear forces. Motivated by the success of the OBE model in describing the $NN$ interaction, Voloshin and Okun applied the OBE model to study the interaction between charmed hadrons forty years ago~\cite{Voloshin:1976ap}. After $X(3872)$ was discovered by the Belle Collaboration in 2003~\cite{Choi:2003ue}, Swanson utilized the one-pion exchange potential to interpret it as a $\bar{D}D^{\ast}$ bound state~\cite{Swanson:2003tb}. Later, Liu et al. studied $X(3872)$ with the OBE model by considering $\sigma$, $\rho$, and $\omega$ exchanges ~\cite{Liu:2008fh,Liu:2008tn}. In Ref.~\cite{Yang:2011wz}  the likely existence of bound states of $\bar{D}^{(\ast)}$ and $\Sigma_{c}$ was predicted in the OBE model (also in the unitary approach~\cite{Wu:2010jy} and quark model~\cite{Wang:2011rga}), which may correspond to the pentaquark states discovered by the LHCb Collaboration ~\cite{Aaij:2015tga,Aaij:2019vzc}. Clearly, the OBE  model has been an invaluable tool in describing hadron-hadron interactions containing not only light quarks but also heavy quarks.

In the conventional OBE model, the interaction between two hadrons  is mainly generated by  exchanging mesons consisting of light up an down quarks.  It is natural to expect that the exchange of mesons containing strange quarks may also play a role in certain systems, such as $K$, $K^{\ast}$, $\eta$, and $\eta^{\prime}$.   In Ref.~\cite{SanchezSanchez:2017xtl}, it was shown that the  one-kaon exchange force is able to bind the $DD_{s0}(2317)$  system.  Compared with the one-pion exchange, $\eta$ and $\eta^{\prime}$ exchanges contribute little  to systems containing only light quarks due to their large masses  and small isospin factor~\cite{Liu:2019zvb}, but they could contribute  more to hadrons containing  strange quarks~\cite{Karliner:2016ith,Wang:2021hql}.
In recent years, to understand the nature of $Z_{c}(3900)$~\cite{BESIII:2013ris,Belle:2013yex}, the one charmonium  exchange has been explored to investigate the $\bar{D}D^{\ast}$ interaction~\cite{Aceti:2014uea,He:2015mja}. For such systems, however, since the exchange of light mesons is also allowed, the one charmonium exchange plays a minor role because of the large masses of charmonia. To confirm the relevance of one charmonium exchange, one needs to study other systems where light meson exchanges are not allowed.

Recently, the HAL QCD Collaboration found one $\Omega_{ccc}\Omega_{ccc}$ dibaryon with a binding energy of $B=5.7$ MeV and a root mean square (RMS) radius of $R=1.1$ fm~\cite{Lyu:2021qsh}, which indicates that there exist strong attractive interactions between the $\Omega_{ccc}\Omega_{ccc}$ pair. In this work we extend the conventional OBE model by allowing for exchanges of  charmonia. We show indeed that the one charmonium exchange can generate attractive interactions strong enough to bind the $\Omega_{ccc}\Omega_{ccc}$ pair. Replacing charm quarks with strange quarks, we find that the one strangeonium exchange leads to the existence of an $\Omega\Omega$ bound state,  consistent with the HAL QCD result~\cite{Gongyo:2017fjb}. Furthermore, we predict  the existence of an $\Omega_{bbb}\Omega_{bbb}$ dibaryon in the extended OBE model with the exchange of  bottomonia.

\begin{figure}[ttt]
\begin{center}
\begin{tabular}{c}
\begin{minipage}[t]{0.7\linewidth}
\begin{center}
\begin{overpic}[scale=.55]{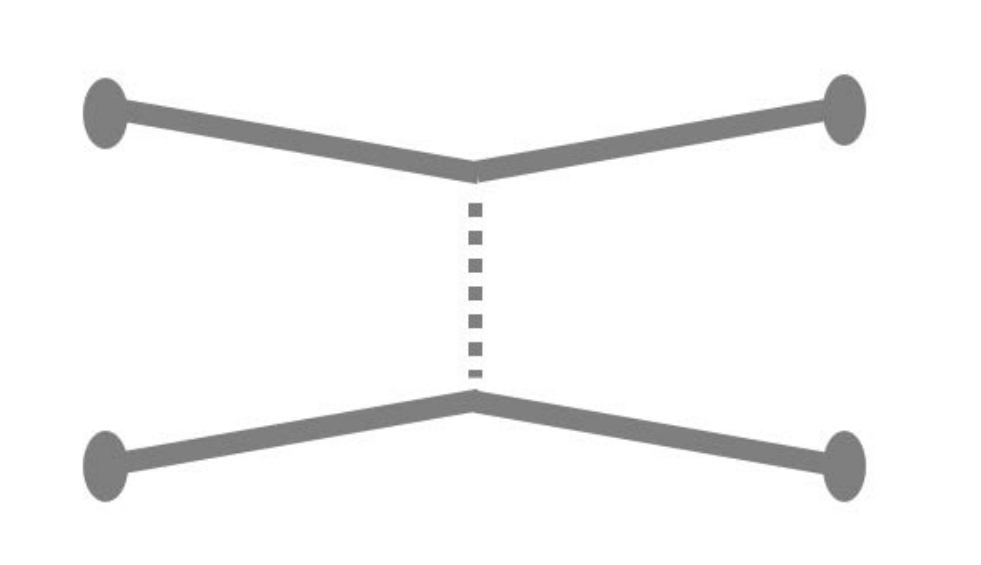}
		\put(89,8){$\Omega_{ccc}$}
		
		\put(-1,9){$\Omega_{ccc}$}
		
		\put(-1,45){$\Omega_{ccc}$}
		\put(89,45){$\Omega_{ccc}$} \put(26,27){$\eta_{c}$, $\chi_{c0}$, $\psi$}
\end{overpic}
\end{center}
\end{minipage}
\end{tabular}
\caption{One charmonium exchange for the $\Omega_{ccc}\Omega_{ccc}$ system  }
\label{obe}
\end{center}
\end{figure}
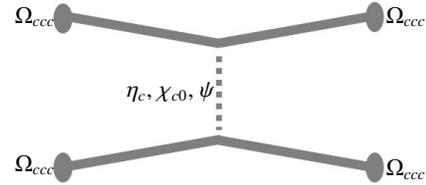
\textit{Theoretical Framework:}
In the OBE  model, the interaction between the $\Omega_{ccc}\Omega_{ccc}$ pair can only be generated by exchanging charmonia, for which we choose the ground-state $\eta_{c}$, $J/\psi$, and $\chi_{c0}(1P)$ mesons. As shown in Fig.~\ref{obe}, they play the same role as the $\pi$, ($\rho$,$\omega$), and $\sigma$  mesons in the conventional OBE model for the $NN$ interaction.   To derive the one charmonium exchange potential of Fig.~\ref{obe}, we need the following Lagrangians~\cite{Lu:2017dvm}:
\begin{eqnarray}
    \mathcal{L}_{\Omega_{ccc}\Omega_{ccc}\eta_{c}} &=& g_{\eta_{c}}
\vec{\Omega}_{ccc}^{\dagger} \cdot(\eta_{c}  \vec{\nabla} \times
 \vec{ \Omega}_{ccc}) \, , \label{eq:L-pi} \\
  \mathcal{L}_{\Omega_{ccc}\Omega_{ccc}\chi_{c0}} &=& g_{\chi_{c0}}
    \vec{\Omega}_{ccc}^{\dagger}\cdot \chi_{c0} \vec{\Omega}_{ccc} \, , \label{eq:L-sigma} \\
  \mathcal{L}_{\Omega_{ccc}\Omega_{ccc}\psi} &=& g_{\psi}
    {\vec{\Omega}_{ccc}}^{\dagger} \cdot \psi \vec{\Omega}_{ccc} \,  \\ \nonumber &-&  \frac{f_{\psi}}{4 M}\vec{\Omega}_{ccc}^{i\dagger}(\partial_{i}\vec{\psi}_{j}-\partial_{j}\vec{\psi}_{i})\vec{\Omega}_{ccc}^{j}, \, \label{eq:L-rho}
\end{eqnarray}
where $\vec{\Omega}_{ccc}$ denotes  the non-relativistic field of $\Omega_{ccc}$,  $\eta_{c}$, $\chi_{c0}$, and $\psi$ are the fields of exchanged mesons,   $g_{\eta_{c}}$, $g_{\chi_{c0}}$, $g_{\psi}$, and $f_{\psi}$  are the corresponding couplings, and the mass scale $M$ renders the coupling $f_{\psi}$ dimensionless.   The $\Omega_{ccc}$  couplings to the charmonia are not known experimentally, but they can be estimated with reasonable confidence as explained below.

In the OBE $NN$ interaction, one can reproduce the binding energy and RMS
radius  of the deuteron with the following nucleon couplings to $\pi$, $\sigma$, and $\omega$: $g_{\pi}=6.84$,  $g_{\sigma }=8.46$,   $g_{v}=3.25$, $f_{v}=19.82$ together with $M=0.94$ GeV, and a cutoff $\Lambda=0.86$ GeV.  Assuming that the OBE $\Omega_{ccc}\Omega_{ccc}$ interaction is similar to the $NN$ interaction by replacing light quarks with charm quarks,  it is reasonable to expect that the couplings $g_{\eta_{c}}$, $g_{\chi_{c0}}$, and $g_{\psi}$ are proportional to  $g_{\pi}$,  $g_{\sigma }$ and   $g_{v}$, respectively. For the sake of simplicity and without the loss of generality, we assume that the ratios are the same, i.e., $g_{\eta_c}/g_\pi=g_{\chi_{c0}}/g_\sigma=g_\psi/g_v=f_\psi/f_v=r$, where $r$ is a free parameter in our model, less than or equal to 1.

\begin{table}[!ttt]
\centering
\caption{Hadron masses (in units of GeV) of relevance  in this work.
}
\label{mass}
\begin{tabular}{c|cc cc c cc}
  \hline \hline
    Hadron  &~~~~  $\Omega_{ccc}$~\cite{Lyu:2021qsh}   &~~~~  $\eta_{c}$ &~~~~ $\chi_{c0}(1P)$  &~~~~ $J/\psi$
     \\  Mass   &~~~~  4.7956 &~~~~ 2.9839 &~~~~ 3.414 &~~~~ 3.096  \\  \hline
        Hadron  &~~~~  $\Omega$  &~~~~  $\eta$ &~~~~ $f_{0}(980)$  &~~~~ $\phi$
     \\  Mass   &~~~~  1.672 &~~~~ 0.958 &~~~~ 0.99 &~~~~ 1.019
     \\  \hline
         Hadron  &~~~~  $\Omega_{bbb}$\cite{Meinel:2010pw}   &~~~~  $\eta_{b}$ &~~~~ $\chi_{b0}(1P)$  &~~~~ $\Upsilon(1S)$
     \\  Mass   &~~~~  14.371 &~~~~ 9.3987 &~~~~ 9.859 &~~~~ 9.460 \\
  \hline \hline
\end{tabular}
\end{table}

As in the conventional OBE model, we have to take into account  the finite size of exchanged mesons by supplementing  the vertices of Fig.~\ref{obe} with a  monopolar form factor
\begin{eqnarray}
F(\Lambda,q)=\frac{\Lambda^2-m^2}{\Lambda^2-q^2}.
\end{eqnarray}
 In the OBE model with only $\pi,\sigma,\rho/\omega$ exchanges, the cutoff $\Lambda$ is the only unknown parameter,  which can be determined by reproducing the binding energies of some molecular candidates. From the physical perspective, the  cutoff should be larger than the masses of exchanged bosons.  Thus, for the one charmonium exchange model,  the cutoff should be larger than the mass of $\chi_{c0}(1P)$. In addition, the cutoff should not be too large to justify the neglect of exchange of heavier mesons.

Using the above Lagrangians,   the one charmonium exchange potentials of the $\Omega_{ccc}\Omega_{ccc}$ system in coordinate space  read
\begin{eqnarray}
  V_{\eta_{c}}(\vec{r}) &=&
    g_{\eta_{c}}^{2}\,\Big[
    - \vec{a}_{1} \cdot \vec{a}_{2}\,\delta(\vec{r})
    + \, \vec{a}_{1} \cdot \vec{a}_{2}\,m_{\eta_{c}}^3\,W_Y(m_{\eta_{c}} r)  \\ \nonumber  &+&S_{12}(\vec{a}_{1},\vec{a}_{2},\vec{r})\,m_{\eta_{c}}^3\, W_{T}(m_{\eta_{c}} r) \Big] \, , \label{11}  \\
  V_{\chi_{c0}}(\vec{r}) &=& -{g_{\chi_{c0} }^2}\,m_{\chi_{c0}}\,W_Y(m_{\chi_{c0}} r)
  \, ,  \label{22} \\
  V_{\psi}(\vec{r}) &=&
    {g_{\psi}^2}\,m_{\psi}\,W_Y(m_{\psi} r) \\ \nonumber &+&\frac{f_{\psi}^{2}}{4M^2}\Big[
    -\frac{2}{3} \vec{a}_{1} \cdot \vec{a}_{2}\,\delta(\vec{r})
    + \,\frac{2}{3} \vec{a}_{1} \cdot \vec{a}_{2}\,m_{\psi}^3\,W_Y(m_{\psi} r) \\ \nonumber
    &+&\frac{1}{3}S_{12}(\vec{a}_{1},\vec{a}_{2},\vec{r})\,m_{\psi}^3\, W_{T}(m_{\psi} r)\Big] \, , \label{33}
\end{eqnarray}
where $\vec{a}_{1} \cdot \vec{a}_{2}$ and $S_{12}(\vec{a}_{1},\vec{a}_{2},\vec{r})$  denote the spin-spin and tensor operators, respectively.  The  functions $W_{Y}(mr)$, $\delta(r)$, and $W_{T}(mr)$  are
\begin{eqnarray}
W_{Y}(mr)&=&\frac{e^{-m r}-e^{-\Lambda r}}{4\pi m r} -
  \frac{\Lambda^2-m^2}{8 \pi \Lambda m}e^{-\Lambda r},
  \\ \nonumber
  \delta(r)&=&
  \frac{(\Lambda^2-m^2)^2}{8 \pi \Lambda}e^{-\Lambda r},  \\ \nonumber
  W_{T}(mr)&=&\frac{3}{4}\frac{e^{-m r}-e^{-\Lambda r}}{\pi m^3 r^3} -\frac{3}{4}\frac{\Lambda e^{-\Lambda r}- m e^{-m r}}{\pi m^3 r^2} \\ \nonumber
  &+&\frac{1}{8\pi m^3r}\Big[(m^2-3\Lambda^2)e^{-\Lambda r}+2 m^2 e^{-m r}\Big]  \\ \nonumber
  &-&\frac{\Lambda^2-m^2}{8\pi m^3}\Lambda e^{-\Lambda r}.
\end{eqnarray}

It is well known that the $S$-$D$ coupling plays an important role in the description of the deuteron, thus we also consider the $S$-$D$ coupling in the extended OBE model.~\footnote{As all the lattice QCD simulations only considered $S$-wave interactions~\cite{Gongyo:2017fjb,Lyu:2021qsh}, in the supplementary material, we discuss the results obtained in the extended OBE model with the $S$-$D$ coupling turned off.} For $J=0$, the allowed orbital angular momenta and spins for the  $\Omega_{ccc}\Omega_{ccc}$ system are $(L=0,S=0)$ and $(L=2,S=2)$, and the relevant matrix elements of the spin-spin and tensors operators are given in Table~\ref{tab:spin1123}.   The $\Omega_{ccc}$ mass is taken from lattice QCD simulations~\cite{Lyu:2021qsh}.  The masses of other relevant charmonia  from PDG~\cite{ParticleDataGroup:2018ovx} are collected in Table~\ref{mass}.

\begin{table}[t]
\centering
\centering \caption{Relevant spin-spin and tensor matrix elements for the $\Omega_{ccc}\Omega_{ccc}$ system.} \label{tab:spin1123}
\begin{tabular}{c|c|c|ccccccc}
\hline\hline
State & $J^{P}$  &   Partial wave    & $\langle \vec{a}_{1}\cdot \vec{a}_{2}\rangle$& $S_{12}(\vec{a}_{1}, \vec{a}_{2}, \vec{r})$ \\ \hline
$\Omega_{ccc}\Omega_{ccc}$  & $J=0$ & $^1S_{0}$-$^5D_{0}$
& $
\left(\begin{matrix}
-\frac{15}{4} & 0 \\
0 & -\frac{3}{4}\\
\end{matrix}\right)$  &$
\left(\begin{matrix}
0 & -3 \\
-3 & -3\\
\end{matrix}\right)$ \\ \hline
 \hline
\end{tabular}
\end{table}

In the same way, the $\Omega\Omega$ and $\Omega_{bbb}\Omega_{bbb}$ potentials can also be constructed  in the extended OBE model. The relevant  particles and their masses~\cite{ParticleDataGroup:2018ovx} are given in  Table~\ref{mass}. The mass of $\Omega_{bbb}$ is also taken from  lattice QCD simulations~\cite{Meinel:2010pw}.

\begin{table}[ttt]
\centering
\caption{Binding energy $B$ and RMS radius $R$ of the $J^{P}=0^{+}$ ${\Omega}_{ccc}{\Omega}_{ccc}$ bound states obtained with different ratios $r$ and cutoffs $\Lambda$. }
\label{dk12}
\begin{tabular}{cc|cccccccc}
\hline\hline
Molecule  & $J^{P}$ ~ & Ratio $r$ ~ & $\Lambda$(GeV) ~ & B (MeV)~  &  R(fm)    \\ \hline
\multirow{2}{0.9cm}{${\Omega}_{ccc}{\Omega}_{ccc}$}  & \multirow{2}{0.2cm}{$0^{+}$} ~ & 1 ~  & 3.78~  &  $5.1$~ & $1.1$
    \\
 &   &1/2 ~  & 4.60~  &  5.8~ & 1.0
    \\
  \hline\hline
\end{tabular}
\end{table}


With the above potentials,
we solve the Schr\"odinger equation to search for bound states.
Assuming that the ratio $r$ is 1 and  with a cutoff  $\Lambda=3.78$ GeV, the binding energy and RMS radius for the $J^{P}=0^{+}$ $\Omega_{ccc}\Omega_{ccc}$ dibaryon is  found to be $B=5.1$ MeV and $R=1.1$ fm, respectively. Decreasing  the ratio to $1/2$ and with a larger cutoff of $\Lambda=4.60$ GeV, the corresponding  binding energy and RMS radius become $B=$ 5.8 MeV and $R=$1.0 fm, respectively. Referring to the cutoff  in the traditional OBE model containing only light quarks, we expect the cutoff in the $\Omega_{ccc}\Omega_{ccc}$ system should be larger than $m_{\chi_{c0}(1P)}=3.414$ GeV, but not too much larger. Therefore, the coupling ratio of $r=1$ is prefered.

From the results shown in Table~\ref{dk12}, we conclude that there exists an $\Omega_{ccc}\Omega_{ccc}$ bound state with charm number $C=6$, corresponding to the dibaryon state discovered by the HAL QCD Collaboration~\cite{Lyu:2021qsh}.
Using  the same couplings and cutoff we did not find an $\Omega_{ccc}\Omega_{ccc}$ bound state with  $J^{P}=2^{+}$ (for more discussions about $J^P=2^+$ configurations, see the supplementary material), consistent with  the quark model study~\cite{Huang:2020bmb}.

Because $\Omega_{ccc}$ has an electric charge  of $+2e$, the repulsive Coulomb  interaction between the $\Omega_{ccc}\Omega_{ccc}$ pair should be taken into account. On top of the one charmonium exchange potential, we add  the $\Omega_{ccc}\Omega_{ccc}$  Coulomb  interaction  $\frac{4\alpha}{r}$.  The updated results show that the  $\Omega_{ccc}\Omega_{ccc}$ system is no longer bound in  the  case of $r=1$, which is in agreement with the  HAL QCD study~\cite{Lyu:2021qsh}. However,  varying the cutoff  by only 0.1 GeV, we find that the  $\Omega_{ccc}\Omega_{ccc}$ system binds with a binding energy of $B=13.8$ MeV. As the reasonable range of the cutoff in the OBE model is  at least  a few hundreds of MeV,  we can not conclude whether the $\Omega_{ccc}\Omega_{ccc}$ system is bound or not once the Coulomb interaction is taken into account. The bottom line is that the Coulomb interaction is important for the $\Omega_{ccc}\Omega_{ccc}$ system.

We can test  the above approach in the $\Omega\Omega$ system, which can exchange $\eta^{\prime}$, $f_{0}(980)$, and $\phi$. The couplings between the triply strange baryon $
\Omega$ and the exchanged mesons are assumed to be proportional to those nucleon couplings as in the one charmonium exchange potential.  The cutoff for the $\Omega\Omega$ system should be greater than the mass of $m_{\phi}=1.019$ GeV.
With the ratio $r=1$ and  cutoff $\Lambda=$1.62 GeV,  we obtain one $J^{P}$=$0^{+}$ $\Omega\Omega$  bound state with a binding energy of $B=$ 1.6 MeV and a RMS radius $r=$ 3.1 fm.
The HAL QCD Collaboration have investigated the $S-$wave interaction of the $\Omega\Omega$ system, and found a bound state with a binding energy of $B=1.6$ MeV  and a RMS radius $r=3\sim4$ fm~\cite{Gongyo:2017fjb}.
If we decrease the ratio $r$ to 1/2, we need a cutoff $\Lambda=2.83$ GeV to reproduce the binding energy and RMS radius of the lattice QCD study~\cite{Gongyo:2017fjb}. In this case, the cutoff value is quite close to $2m_{\phi}$, which does not seem very natural. This  then implies that the ratio in the strangeness sector should be greater than 1/2. We note that allowing for mixing of light quarks with strange quarks, a  larger coupling ratio than that of the charm sector is expected.
Compared with the $\Omega_{ccc}\Omega_{ccc}$ dibaryon, the $\Omega\Omega$ dibaryon has a larger size, which can be easily understood because on the one hand the size of $\Omega$ is larger than  that of $\Omega_{ccc}$ and on the other hand the exchanged mesons are lighter and therefore the interaction range is longer.

From the above study, it is clear that the extended OBE model can be trusted to study baryon-baryon interactions containing only heavy quarks such as charm and strangeness. As a result, it is natural to  ask how about the $\Omega_{bbb}\Omega_{bbb}$ system, which can exchange $\eta_{b}$, $\chi_{b0}(1P)$, and $\Upsilon(1S)$. The couplings of the triply bottom baryon $\Omega_{bbb}$ and bottomia  can  also be estimated from the couplings of the nucleon with light mesons.
A reasonable cutoff for the $\Omega_{bbb}\Omega_{bbb}$ system is expected to be larger than the mass of $m{\chi_{b0}(1P)}=9.859$ GeV. With a cutoff $\Lambda=10.073$ GeV and the corresponding ratio $r=1$  we obtain one $J^P=0^+$ bound state with a binding energy $B= 5.7$ MeV and a RMS radius $R=0.55$ fm. If we decrease the ratio to 1/2 and increase the cutoff to $\Lambda=10.718$ GeV, the corresponding binding energy and  RMS radius become $B=$5.7 MeV and $R=0.55$ fm, respectively. Compared with the sizes of  the $\Omega_{ccc}\Omega_{ccc}$ and $\Omega\Omega$ dibaryons, the size of the $\Omega_{bbb}\Omega_{bbb}$ dibaryon is  small. Nonetheless, one can see that the ratio of the size of the $\Omega_{bbb}\Omega_{bbb}$ dibaryon to the size of $\Omega_{bbb}$ is similar to the ratio of their charm counterparts, $R_{\Omega_{ccc}\Omega_{ccc}}$/$R_{\Omega_{ccc}}$.

  Both $\Omega$  and $\Omega_{bbb}$  contain one negative charge  $-e$, which leads to a repulsive potential $\frac{\alpha}{r}$ between the $\Omega\Omega$ and $\Omega_{bbb}\Omega_{bbb}$ pairs.  With the Coulomb potential taken into account we update the results in Table~\ref{dk3}. The numbers in the parentheses show that the  $\Omega\Omega$ and  $\Omega_{bbb}\Omega_{bbb}$ systems  still bind, though less bound than in the case where only the OBE potentials are considered. In addition, we find  that the impact of the Coulomb interaction on the $\Omega\Omega$ system is smaller than that on the $\Omega_{bbb}\Omega_{bbb}$ system,  mainly because the  $\Omega\Omega$ dibaryon is more extended than the    $\Omega_{bbb}\Omega_{bbb}$ dibaryon.

\begin{table}[ttt]
\centering
\caption{Binding energies $B$ and RMS radii $R$ of $J^{P}=0^{+}$ ${\Omega}{\Omega}$ and  $\Omega_{bbb}\Omega_{bbb}$ bound states obtained with different ratio $r$ and cutoff $\Lambda$. The numbers in the brackets represent the results obtained with  the Coulomb interaction taken into account.   }
\label{dk3}
\begin{tabular}{cc|cccccccc}
\hline\hline
Molecule ~ & $J^{P}$ ~  & Ratio r~    & $\Lambda$(GeV)~   & B(MeV) ~~  &  R(fm)   \\
\hline
\multirow{2}{0.5cm}{$\Omega\Omega$} ~ & \multirow{2}{0.2cm}{$0^{+}$}  ~ & 1 ~  & 1.62~ &  $1.6(\textbf{0.7})$ ~~& 3.1(\textbf{4.0})
    \\
  &   &    1/2 ~ & 2.83~ & $1.6(\textbf{0.6})$~~ & $3.0(\textbf{4.2})$
    \\  \hline
  \multirow{2}{1.0cm}{${\Omega_{bbb}}{\Omega_{bbb}}$}  & \multirow{2}{0.2cm}{$0^{+}$} & 1  & 10.073  &  $5.7$(\textbf{0.5}) & 0.55(\textbf{1.12})
    \\
         &  &   1/2  & 10.718 & $5.7$(\textbf{0.4}) & 0.55(\textbf{1.17})
    \\
  \hline\hline
\end{tabular}
\end{table}

\begin{figure}[!h]
\begin{overpic}[scale=.33]{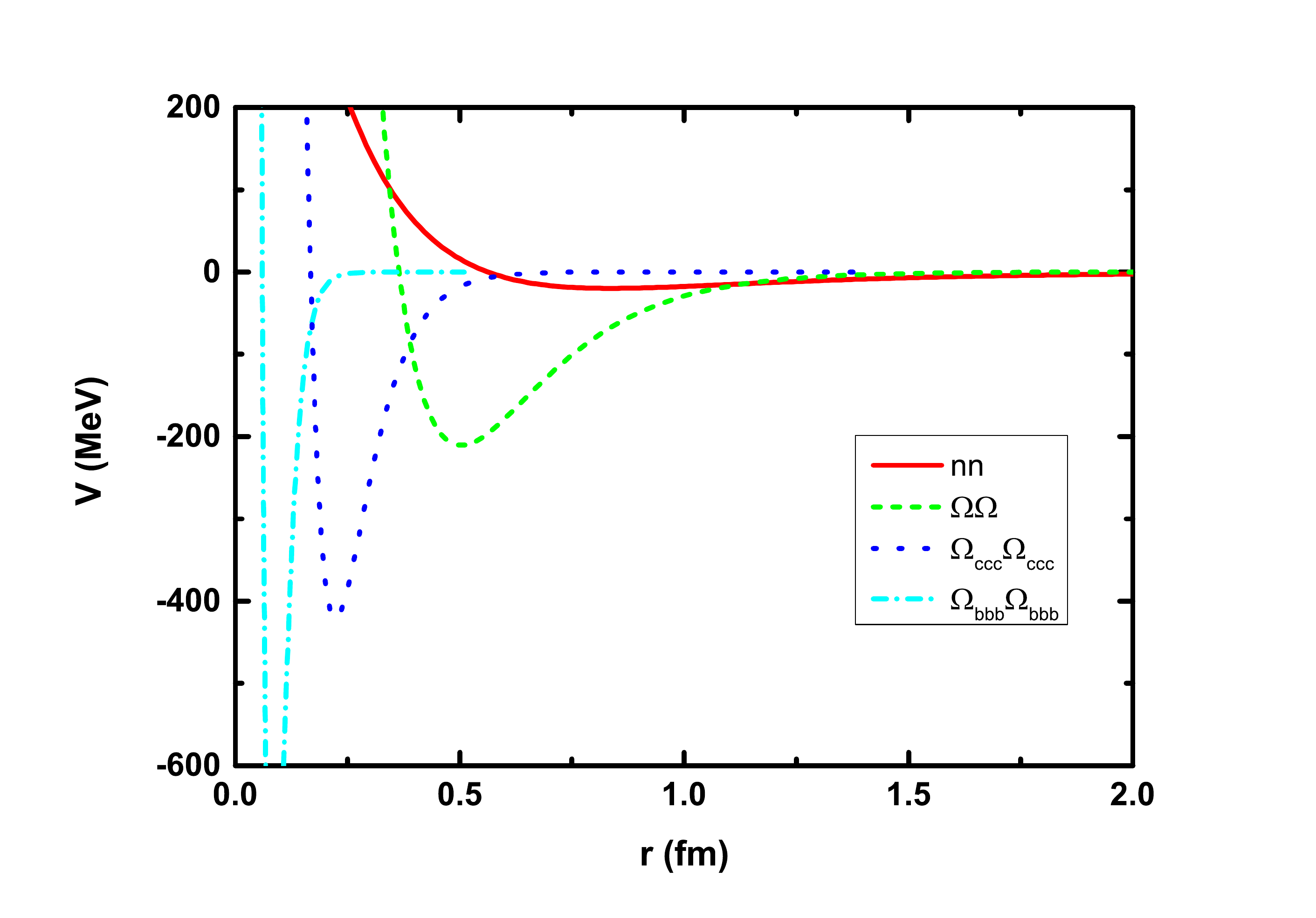}
\end{overpic}
\caption{One boson exchange potential for the $NN$, $\Omega\Omega$, $\Omega_{ccc}\Omega_{ccc}$, and $\Omega_{bbb}$-$\Omega_{bbb}$ systems as functions of the distance between the baryon-baryon pair.    }
\label{pot}
\end{figure}

With the coupling ratio $r=1$ and the relevant cutoffs given in Tables~\ref{dk12} and ~\ref{dk3}, we show  the $S$-wave potential of $NN$, $\Omega\Omega$, $\Omega_{ccc}\Omega_{ccc}$, and $\Omega_{bbb}\Omega_{bbb}$ in Fig.~\ref{pot}.  The line shapes of these potentials are similar, i.e., they all contain a short-range repulsive core and a medium- and long-range  attractive tail. As the baryon mass becomes heavier, the baryon-baryon potential becomes more short ranged and attractive, which would generate a delta-like potential in the limit of infinite heavy baryon masses.  We note that  the $^3S_{1}$  $NN$ potential is qualitatively consistent with those of  more refined phenomenological models~\cite{Machleidt:2000ge,Stoks:1994wp} as well as  lattice QCD simulations~\cite{Ishii:2006ec}.

\begin{figure}[ttt]
\begin{center}
\begin{tabular}{c}
\begin{minipage}[t]{0.9\linewidth}
\begin{center}
\begin{overpic}[scale=.7]{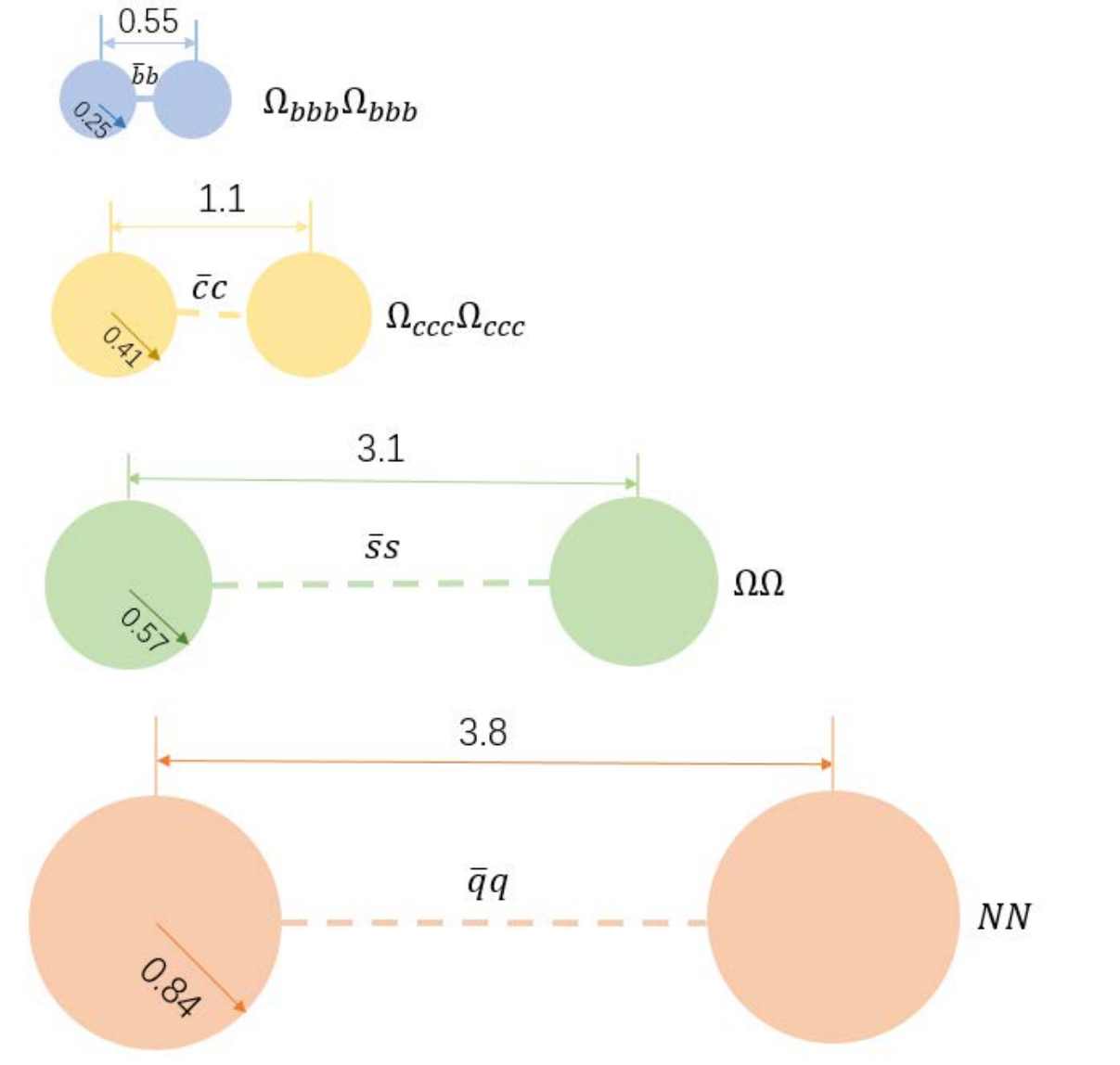}
\put(60,78){Units:~\textbf{fm}}
\end{overpic}
\end{center}
\end{minipage}
\end{tabular}
\caption{ RMS radii of the dibaryons and their components. Those of the nucleon, $\Omega$, $\Omega_{ccc}$, and $\Omega_{bbb}$  are 0.84 fm~\cite{ParticleDataGroup:2018ovx}, 0.57 fm~\cite{Can:2015exa}, 0.41 fm~\cite{Can:2015exa}, 0.25 fm~\cite{Meinel:2010pw}, while those of the $NN$, $\Omega\Omega$, $\Omega_{ccc}\Omega_{ccc}$, and $\Omega_{bbb}\Omega_{bbb}$ dibaryons calculated in the extended OBE model are 3.8 fm, 3.1 fm, 1.1 fm, and 0.55 fm, respectively, with the ratio $r=1$ and the cutoff $\Lambda=0.86,1.62,3.78,10.073$ GeV, respectively.   }
\label{size}
\end{center}
\end{figure}

In Fig.~\ref{size}, we show the RMS radii of deuteron, $\Omega\Omega$, $\Omega_{ccc}\Omega_{ccc}$, and $\Omega_{bbb}\Omega_{bbb}$ dibaryon states calculated in the extended OBE model as well as  those of their components~\footnote{We do not distinguish charge radius from matter radius, which are assumed to be equal or similar for baryons/dibaryons studied here.}. It is clear that for all the dibaryons studied, the radius of the dibaryon is larger than  the sum of those of their components. As a result, it is reasonable to refer to these dibaryons as molecular states.  Moreover,  the RMS radius of the dibaryon state becomes  smaller as the baryon mass increases, because the exchanged mesons become heavier. The ratio of the size of the dibaryon to its component  $R_{BB}/R_B$ is 4.5, 5.4, 2.7, 2.2, for the $NN$, $\Omega\Omega$, $\Omega_{ccc}\Omega_{ccc}$, and $\Omega_{bbb}\Omega_{bbb}$ dibaryons, respectively. These results exhibit approximate SU(3) symmetry and  heavy quark flavor symmetry, as expected.


It should be noted that the results obtained in the extended OBE model are cutoff dependent, as in the conventional OBE model. In fixing the cutoffs for the $\Omega_{ccc}\Omega_{ccc}$ and $\Omega\Omega$ systems, we have used the lattice QCD simulations as guidance. While for the $\Omega_{bbb}\Omega_{bbb}$ system, we have used the binding energy of the $\Omega_{ccc}\Omega_{ccc}$ dibaryon as a reference. The resulting cutoffs all seem very reasonable, which strongly support the existence of $\Omega\Omega$, $\Omega_{ccc}\Omega_{ccc}$, and $\Omega_{bbb}\Omega_{bbb}$ dibaryons. In particular, for the $\Omega_{bbb}\Omega_{bbb}$ system, a naive application of heavy quark flavor symmetry to connect $\Omega_{ccc}\Omega_{ccc}$ to $\Omega_{bbb}\Omega_{bbb}$ would result in a binding energy of a few tens of MeV for the latter, which implies that the Coulomb interaction will not likely break up the $\Omega_{bbb}\Omega_{bbb}$ pair, further supporting our conclusion.

 At last, we briefly discuss about the $B\bar{B}$ system using the $\Omega_{ccc}\bar{\Omega}_{ccc}$ system as an example. If only  the Coulomb interaction is considered,   we obtain two bound states with binding energies $B=$1.0 MeV and $B=$0.2 MeV. The one charmonium exchange potential for the  $\Omega_{ccc}\bar{\Omega}_{ccc}$ system  is related  to  that for the  $\Omega_{ccc}\Omega_{ccc}$ system  via $C$-parity, analogous with the $G$-parity  in the $NN$ system~\cite{Klempt:2002ap}. Therefore, the only difference between the $\Omega_{ccc}\bar{\Omega}_{ccc}$ potential and  the $\Omega_{ccc}\Omega_{ccc}$ potential comes from the one $J/\psi$ exchange potential. With the same parameters as those used for the  $\Omega_{ccc}\Omega_{ccc}$ system, we obtain  a deeply bound $\Omega_{ccc}\bar{\Omega}_{ccc}$ state, because the $\Omega_{ccc}\bar{\Omega}_{ccc}$ potential is more  attractive, similar to the  nucleon–antinucleon case~\cite{Klempt:2002ap}.  It is natural to expect that the   $\Omega_{ccc}\bar{\Omega}_{ccc}$ system may generate a bound state  after considering both the Coulomb potential  and the one charmonium exchange potential.

\textit{Summary and outlook:} Replacing light $u/d$ quarks with heavy (strangeness, charm, and bottom) quarks, we extended the conventional OBE model by allowing for exchanges of heavy $s\bar{s}$, $c\bar{c}$, $b\bar{b}$ ground-state mesons. Fixing the nucleon couplings to light mensons  by  reproducing the binding energy and RMS radius of the deuteron, and assuming that the $\Omega$, $\Omega_{ccc}$, and $\Omega_{bbb}$ couplings to their relevant exchanged  mesons are equivalent to or less than the nucleon ones,  we investigated the likely existence of   $\Omega\Omega$, $\Omega_{ccc}\Omega_{ccc}$, and $\Omega_{bbb}\Omega_{bbb}$ dibaryons via the extended OBE model. Our results confirmed the existence of $J^{P}=0^{+}$ $\Omega\Omega$ and  $\Omega_{ccc}\Omega_{ccc}$ bound states found by lattice QCD simulations. Furthermore, we  predicted the existence of  one $J^{P}=0^{+}$ $\Omega_{bbb}\Omega_{bbb}$ bound state.
In addition, taking into account the Coulomb interaction, we found that the  $\Omega_{ccc}\Omega_{ccc}$ system is at the verge of binding, while the $\Omega\Omega$ and $\Omega_{bbb}\Omega_{bbb}$ systems still bind.  Our studies showed that  the spin-parities of such dibaryons favor $J^{P}=0^{+}$ than $J^{P}=2^{+}$.

\textit{Acknowledgements:} This work was partly supported by the National Natural Science Foundation of China (NSFC) under Grants No. 11975041, No.11735003, and No.11961141004.

\section{Supplementary material}
In this supplementary material we provide additional information not detailed
in the main text which might help the reader to better understand
the results presented in this work.

\subsection{Only $S$-wave interaction }

To directly compare with the lattice QCD studies~\cite{Gongyo:2017fjb,Lyu:2021qsh},  we show the results obtained with  only $S$-wave interactions in the extended OBE model. Turning off the $S$-$D$ mixing, but with the same cutoffs $\Lambda$ and coupling ratios $r$ as given in Tables  III and IV of the main text, we repeat the OBE study of the $\Omega_{ccc}\Omega_{ccc}$, $\Omega\Omega$, and $\Omega_{bbb}\Omega_{bbb}$ systems and  search for bound states. The  results  are shown in Table~\ref{dk12}. It is clear that without the $S$-$D$ mixing, all the systems become less bound. Adding the Coulomb interaction, some of the dibaryons even become unbound. On the other hand, with a slight fine-tuning (increase) of the cutoff or coupling ratio, the systems can become bound as shown in Table~\ref{s0}. Therefore, as concluded in the main text, the extended OBE model supports the main results of the lattice QCD studies~\cite{Gongyo:2017fjb,Lyu:2021qsh}.
\begin{table}[!h]
\centering
\caption{Binding energy $B$ and RMS radius $R$ of the $J^{P}=0^{+}$ $\Omega_{ccc}\Omega_{ccc}$, $\Omega\Omega$ and $\Omega_{bbb}\Omega_{bbb}$ bound states obtained with different ratios $r$ and cutoffs $\Lambda$. The numbers in the brackets represent the results obtained with  the Coulomb interaction taken into account. The ``$-$'' denotes that there exists no bound state.}
\label{dk12}
\begin{tabular}{cc|cccccccc}
\hline\hline
Molecule  & $J^{P}$ ~ & Ratio $r$ ~ & $\Lambda$(GeV) ~ & B (MeV)~  &  R(fm)    \\ \hline
\multirow{2}{0.9cm}{${\Omega}_{ccc}{\Omega}_{ccc}$}  & \multirow{2}{0.2cm}{$0^{+}$} ~ & 1 ~  & 3.78~  &  $2.6(-)$~ & $1.4(-)$
    \\
 &   &1/2 ~  & 4.60~  &  $1.9(-)$~ & $1.6(-)$
    \\  \hline \multirow{2}{0.5cm}{$\Omega\Omega$} ~ & \multirow{2}{0.2cm}{$0^{+}$}  ~ & 1 ~  & 1.62~ &  $0.9(\textbf{0.2})$ ~~& 4.0(\textbf{6.1})
    \\
  &   &    1/2 ~ & 2.83~ & $0.3(-)$~~ & $5.9(-)$
    \\  \hline
  \multirow{2}{1.0cm}{${\Omega_{bbb}}{\Omega_{bbb}}$}  & \multirow{2}{0.2cm}{$0^{+}$} & 1  & 10.073  &  $1.1(-)$ & $1.2(-)$
    \\
         &  &   1/2  & 10.718  &  $0.5(-)$ & $1.7(-)$
    \\
  \hline\hline
\end{tabular}
\end{table}

\begin{table}[!h]
\centering
\caption{Binding energy $B$ and RMS radius $R$ of the $J^{P}=0^{+}$ $\Omega_{ccc}\Omega_{ccc}$, $\Omega\Omega$ and $\Omega_{bbb}\Omega_{bbb}$ bound states obtained with different ratios $r$ and cutoffs $\Lambda$. The numbers in the brackets represent the results obtained with  the Coulomb interaction taken into account. The ``$-$'' denotes that there exists no bound state.}
\label{s0}
\begin{tabular}{cc|cccccccc}
\hline\hline
Molecule  & $J^{P}$ ~ & Ratio $r$ ~ & $\Lambda$(GeV) ~ & B (MeV)~  &  R(fm)    \\ \hline
\multirow{2}{0.9cm}{${\Omega}_{ccc}{\Omega}_{ccc}$}  & \multirow{2}{0.2cm}{$0^{+}$} ~ & 1 ~  & 3.8075~  &  $5.7(-)$~ & $1.0(-)$
    \\
 &   &1/2 ~  & 4.681~  &  5.7(-)~ & 1.0(-)
    \\  \hline \multirow{2}{0.5cm}{$\Omega\Omega$} ~ & \multirow{2}{0.2cm}{$0^{+}$}  ~ & 1 ~  & 1.64~ &  $1.6(\textbf{0.7})$ ~~& 3.2(\textbf{4.1})
    \\
  &   &    1/2 ~ & 2.98~ & $1.6(\textbf{0.6})$~~ & $3.0(\textbf{4.1})$
    \\  \hline
  \multirow{2}{1.0cm}{${\Omega_{bbb}}{\Omega_{bbb}}$}  & \multirow{2}{0.2cm}{$0^{+}$} & 1  & 10.10  &  $5.3$(\textbf{0.3}) & 0.57(\textbf{1.3})
    \\
         &  &   1/2  & 10.79 & $5.7$(\textbf{0.5}) & 0.55(\textbf{1.1})
    \\
  \hline\hline
\end{tabular}
\end{table}

\subsection{  $J=2$ configurations}
In the following, we discuss about the $J=2$ configurations for the $\Omega\Omega$, $\Omega_{ccc}\Omega_{ccc}$, and $\Omega_{bbb}\Omega_{bbb}$ systems.
For  $J=2$, the following combinations of  angular momentum and spin are allowed: $(L=0,S=2)$, $(L=2,S=0)$, and $(L=2,S=2)$. With these partial waves the relevant matrix elements of the spin-spin  and tensor operators of the extended OBE potential are shown in Table~\ref{tab:spin456}.

\begin{table}[!h]
\centering
\centering \caption{Matrix elements of the spin-spin and tensor operators for the $\Omega_{ccc}\Omega_{ccc}$ system.} \label{tab:spin456}
\begin{tabular}{c|c|c|ccccccc}
\hline\hline
State & $J^{P}$  &   Partial wave    & $\langle \vec{a}_{1}\cdot \vec{a}_{2}\rangle$& $S_{12}(\vec{a}_{1}, \vec{a}_{2}, \vec{r})$ \\ \hline
$\Omega_{ccc}\Omega_{ccc}$  & $J=2$ & $^5S_{2}$-$^1D_{2}$-$^5D_{2}$   & $
\left(\begin{matrix}
-\frac{3}{4} & 0  &0 \\
0 & -\frac{15}{4} & 0  \\
0&  0&  -\frac{3}{4}\\
\end{matrix}\right)$  &$
\left(\begin{matrix}
0 & -\frac{3}{\sqrt{5}}   &3\sqrt{\frac{7}{10}} \\
-\frac{3}{\sqrt{5}}  &0  & 3\sqrt{\frac{2}{7}}  \\
3\sqrt{\frac{7}{10}} & 3\sqrt{\frac{2}{7}}&  \frac{9}{14} \\
\end{matrix}\right)$ \\ \hline
 \hline
\end{tabular}
\end{table}

First, we notice that with the same parameters as those used for the $J=0$ configurations studied in the main text, none of these systems bind, which indicates that in the extended OBE model, the $J=0$ potential is more attractive than that of $J=2$, in agreement with the quark model~\cite{Huang:2020bmb}.

Next we  investigate whether these systems can bind with reasonable cutoffs.
For the $\Omega_{ccc}\Omega_{ccc}$ system,
with a cutoff of 4.18 GeV and  the coupling ratio $r=1$, we obtain one bound state. If we change the coupling ratio to $1/2$, we obtain a bound state for a cutoff of 5.45 GeV. Both cutoffs are much larger than those for $J=0$ by about 0.4-0.9 GeV, which do  not seem very natural. With the same approach, we obtain the results for the $\Omega\Omega$ and $\Omega_{bbb}\Omega_{bbb}$  as shown in Table~\ref{dksd2}. Judging from the cutoffs needed to obtain bound states, we conclude that although a $J=2$ $\Omega_{bbb}\Omega_{bbb}$ dibaryon is unlikely, a $J=2$ $\Omega\Omega$ dibaryon cannot be ruled out.

\begin{table}[!h]
\centering
\caption{Binding energy $B$ and RMS radius $R$ of the $J^{P}=2^{+}$ $\Omega_{ccc}\Omega_{ccc}$, $\Omega\Omega$ and $\Omega_{bbb}\Omega_{bbb}$ bound states obtained with different ratios $r$ and cutoffs $\Lambda$. The numbers in the brackets represent the results obtained with  the Coulomb interaction taken into account. The `` $-$" denotes that there exists no bound state.}
\label{dksd2}
\begin{tabular}{cc|cccccccc}
\hline\hline
Molecule  & $J^{P}$ ~ & Ratio $r$ ~ & $\Lambda$(GeV) ~ & B (MeV)~  &  R(fm)    \\ \hline
\multirow{2}{0.9cm}{${\Omega}_{ccc}{\Omega}_{ccc}$}  & \multirow{2}{0.2cm}{$2^{+}$} ~ & 1 ~  & 4.18~  &  $7.6(-)$~ & $0.9(-)$
    \\
 &   &1/2 ~  & 5.45~  &  $6.6(-)$~ & $0.9(-)$
    \\  \hline \multirow{2}{0.5cm}{$\Omega\Omega$} ~ & \multirow{2}{0.2cm}{$2^{+}$}  ~ & 1 ~  & 1.80~ &  $5.4(\textbf{3.9})$ ~~& 1.8(\textbf{2.0})
    \\
  &   &    1/2 ~ & 3.83~ & $2.5(\textbf{1.1})$~~ & $2.4(\textbf{3.1})$
    \\  \hline
  \multirow{2}{1.0cm}{${\Omega_{bbb}}{\Omega_{bbb}}$}  & \multirow{2}{0.2cm}{$2^{+}$} & 1  & 10.393  &  $3.5(-)$ & $0.67(-)$
    \\
         &  &   1/2  & 11.378 & $5.3(-)$ & $0.55(-)$
    \\
  \hline\hline
\end{tabular}
\end{table}

\bibliography{omegaccc-dibaryon}

\end{document}